\documentclass[preprint,aps]{revtex4}
\usepackage{amssymb}
\usepackage{amsmath}
\usepackage{graphicx}% Include figure files
\usepackage{dcolumn}% Align table columns on decimal point
\usepackage{bm}% bold math

\begin{document}

%\preprint{APS/123-QED}

\title{Kink-Based Path Integral Calculations of Atoms He-Ne }% Force line breaks with \\

\author{Randall W. Hall}
 \altaffiliation[Also at ]{Department of Physics and Astronomy, Louisiana State University}%Lines break automatically or can be forced with \\
\affiliation{Department of Chemistry \\ Louisiana State
University\\Baton Rouge, La. 70803-1804} \email{rhall@lsu.edu}
 \homepage{http://chemistry.lsu.edu/chem/facultypages/hall/hall.asp}

\date{\today}% It is always \today, today,

\begin{abstract}
An adaptive, kink-based path integral formalism is used to
calculate the ground state energies of the atoms He-Ne.  The
method uses an adaptive scheme to virtually eliminate the sign
difficulties.  This is done by using a Monte Carlo scheme to
identify states that contribute significantly  to the canonical
partition function and then include them in the wavefunctions to
calculate the canonical averages. The calculations use the 6-31G
basis set and obtain both precision and accuracy.
\end{abstract}

\pacs{02.70.Rr , 05.30.Fk,31.10.+z,31.15.Gy,
31.25.Eb}% PACS, the Physics and Astronomy
\maketitle

\section{\label{sec:level1}Introduction}
The path integral formulation of quantum mechanics offers a
variety of advantages for studying the electronic and geometric
structures of multi-electron systems\cite{fandh}. Chief among
these are inclusion of finite temperatures (particularly as they
affect geometric degrees of freedom) and exact inclusion of
electron-electron correlation. The application of this method to
electronic systems has been hindered by the so-called "sign"
problem, which results from sign of the fermion density matrix,
which can be positive or negative and leads to large uncertainties
in quantities evaluated using statistical methods such as Monte
Carlo
simulations\cite{piMak1,piMak2,piMak3,piOkazaki,ceperleypiH,kukipi,rwhpihub,rwhpiks,ceperley3he,voth1,voth2,ceperleyh,ceperleyfixednode,AFMC1,AFMC2,AFMC3,AFMC4,pikink1}.
We have recently introduced a "kink-based" path integral
approach\cite{pikink1}, which was demonstrated to overcome the
sign problem in the 2-D Hubbard model. This approach is
complimentary to the  shifted-contour auxiliary-field Monte Carlo
method\cite{AFMC1,AFMC2,AFMC3,AFMC4}, which uses the
Hubbard-Stratonovich transformation to combat the sign problem. In
this work, we use the kink-based formalism to study atomic
systems, the next step in studying systems with  geometric degrees
of freedom (such as atomic clusters).
\section{\label{sec:level2}Kink-Based approach}
In this section, a brief review of the kink-based
approach\cite{pikink1} is given, with additional attention given
to the different spin states that are encountered in electronic
systems. The partition function is written:
\begin{eqnarray}
Q&=&Tr\{\exp(-\beta H)\}\nonumber \\
&=&\sum_{\sigma,\alpha}<\alpha,\sigma|\exp(-\beta H)|\ \nonumber
\alpha,\sigma>\\
&=&\sum_{\sigma,\alpha}\exp(-\beta E_{\alpha,\sigma})
\end{eqnarray}
where $\alpha$ labels the different electronic states associated
with a particular spin state $\sigma$ and $|\ \alpha,\sigma>$ is
the properly anti-symmetrized state. For large enough $\beta$,
this becomes
\begin{eqnarray}
Q\approx\exp(-\beta E_{0,\sigma^*})
\end{eqnarray}
where $E_{0,\sigma^*}$ is the ground state energy of the lowest
energy spin-state. If an approximate set of states, $\{a,s\}$ is
used, we have
\begin{eqnarray}
Q_{\{a,s\}}&=&\sum_{a,s}<a,s|\exp(-\beta H)|\ a,s>\nonumber\\
&=&\sum_{a,s}\sum_{\sigma,\alpha}|<a,s|\alpha,\sigma>|^2
\exp(-\beta E_{\alpha,\sigma})
\end{eqnarray}
As long as  $<a,s|0,\sigma^* >\ \neq 0$ for some $a$ and $s$, then
as $\beta$ gets large,
\begin{eqnarray}
Q_{\{a,s\}}&\propto&\exp(-\beta E_{0,\sigma^*})
\end{eqnarray}
In a later section, we will choose our states with specific values
of $S_z $. Consequently, we will determine the low temperature
partition function corresponding to the lowest energy spin-state
$S$ that has  $S_z $ as one of its possible values of $\hat{S}_z
$.

To evaluate the partition function $Q_{\{a,s\}}$ using the path
integral method, we insert complete sets of states in order to use
the high temperature, semi-classical approximation for the density
matrix:
\begin{eqnarray}
Q_{\{a,s\}}&=&\sum_{a_1 ,s_1 }\cdots  \sum_{a_P ,s_P }<a_1 ,s_1
|\exp(-\beta H/P)|a_2 ,s_2
>\cdots \nonumber \\
&&\times <a_P ,s_P |\exp(-\beta H/P)|a_1
,s_1 >\nonumber\\
&\equiv&\sum_{a_1 ,s_1 }\cdots\sum_{a_P ,s_P }t_{a_1 ,s_1 }^{a_2
,s_2 }\cdots t_{a_P ,s_P }^{a_1 ,s_1 }\label{Qas}
\end{eqnarray}
We refer to a matrix element $<a,s|\exp(-\beta H/P)|a',s'>$ with
$a\neq a'$ or $s\neq s'$ as a \textit{kink}. We rewrite the
partition function as a sum over kinks:
\begin{eqnarray}
Q_{\{a,s\}}&=&\sum_{a,s}\left(t_{a,s}^{a,s}\right)^P + \nonumber \\
&&\sum_{i=1}^P \sum_{a,s}\sum_{a',s'} \left(t_{a,s}^{a,s}\right)^i
\left(t_{a',s'}^{a',s'}\right)^{P-2+i}
\left(t_{a,s}^{a',s'}\right)^2 +\cdots\\
&\equiv&Q_0 + Q_2 + Q_3 + \cdots+Q_P
\end{eqnarray}
where $Q_n $ is the partition function corresponding to $n$ kinks.
 In our
previous work, we demonstrated that $Q_{\{a,s\}}$ has the form
\begin{eqnarray}
Q_{\{a,s\}}&=&\sum_{j}x_{j}^{P}+  \nonumber \\
&&\sum_{n=2}^{P}\frac{P}{n}\left(
\prod_{i=1}^{n}\sum_{j_{i}}\right) \left(
\prod_{k=1}^{n}t_{j_{k},j_{k+1}}\right) S\left( \left\{
x_{j}\right\} ,n,m,\left\{ g_{j}\right\} \right) \label{finaleqn}
\end{eqnarray}
where
\begin{eqnarray*}
x_j =<\alpha_j ,s_j |\exp(-\beta H/P)|\ \alpha_j ,s_j >\approx <\alpha_j ,s_j |(1-\beta H/P)|\ \alpha_j ,s_j >\\
t_{j,j'} =<\alpha_j ,s_j |\exp(-\beta H/P)|\
\alpha_{j'},s_{j'}>\approx <\alpha_j ,s_j |(1-\beta H/P)|\
\alpha_{j'},s_{j'}>
\end{eqnarray*}
and $S\left( \left\{ x_{j}\right\} ,n,m,\left\{ g_{j}\right\}
\right)$ is the contribution to the partition function with $n$
kinks, comprised of $m$ states $\alpha_j $, each occurring $g_j $
times ($\sum_j g_j = P-n$). The explicit form for $S$ is
\begin{eqnarray}
S\left( \left\{ x_{j}\right\} ,n,m,\left\{ g_{j}\right\} \right)
&=&\sum_{l=0}^{m}\frac{1}{(g_l -1)!}\frac{d^{g_l -1}}{dx_{l}^{g_l
-1}}\frac{x_{l}^{P-1 }}{\prod_{k\neq l}(x_l - x_k )^{g_k }}
\label{Final Result for S}
\end{eqnarray}
The derivatives can be evaluated recursively. If we define
\begin{eqnarray}
F^{(p)}_l
&\equiv&\frac{d^{p}}{dx_{l}^{p}}\frac{x_{l}^{P-1}}{\prod_{k\neq
l}\left( x_{l}-x_{k}\right)^{g_k } }
\end{eqnarray}
we can show
\begin{eqnarray}
S&=&\sum_{l=1}^{m}\frac{F_{l}^{(g_l -1)}}{(g_l -1)!}\\
F_{l}^{(n)}&=&\sum_{m=0}^{n-1}\binom{n-1}{m}G_{l}^{(m)}F_{l}^{(n-1-m)}\\
G_{l}^{(m)}&=&(-1)^m m! \left[\frac{P-1}{x_{l}^{m+1}}-\sum_{k\neq
l}\frac{g_k }{(x_l - x_k )^{m+1}}\right]\\
\end{eqnarray}
A similar manipulation leads to an expression for the energy
estimator:
\begin{eqnarray}
E_{est}&=&\sum_{i=1}^{n}\frac{t'_{i,i+1}}{t_{i,i+1}}+ \nonumber \\
&&\frac{1}{S}\sum_{l=0}^{m}\frac{1}{(g_l -1)!}\sum_{j=0}^{g_l
-2}\binom{g_l -2}{j}\left[D_{l}^{(j)}F_{l}^{(g_l
-2-j)}+G_{l}^{(j)}E_{l}^{(g_l -2-j)}\right]\\
E_{l}^{(m)}&\equiv&-\frac{d}{d\beta}F_{l}^{(m)}\\
D_{l}^{(m)}&\equiv&-\frac{d}{d\beta}G_{l}^{(m)}
\end{eqnarray}

The expression shown in Eqn.~\ref{Final Result for S} gives the
exact value of Q (including electron-electron correlation) within
the approximations inherent in using a finite basis set and a
finite level of discretization. The so-called "sign problem" can
occur in this and any other discretized version of the path
integral problem because any of the matrix elements $t_{i,i+1}$
can be negative, resulting in a large variances when evaluating
the partition function using simulation methods. Our approach to
minimizing or eliminating the sign problem has been to apply
Eqn.~\ref{Final Result for S} in an adaptive manner. We first
realize that  the zero kink contribution to Q has no sign
problems, since the system is in a single state. With a properly
chosen states, Q can be obtained with just a few kinks; this
 significantly reduces the sign problem by reducing the statistical error from
greater than 100\% to a precision adequate for chemical
applications. A good choice
 for the states is obtained using a Monte Carlo
 simulation, in which the different N-electron states that appear
 during the simulation are used to update the estimates of the
 ground and excited states. We call
this approach an adaptive approach, since the Monte Carlo
algorithm allows the estimates for the ground and excited state
wavefunctions to evolve according to the statistical sampling of
the different N-electron states.

We implemented the adaptive scheme in the following way (other
methods of are possible). An initial set of basis functions (the
6-31G basis set in our calculations) was orthonormalized and used
to create a set of one-electron orbitals. The one-electron
Hamiltonian was then diagonalized in this basis. Each electron was
assigned a spin and the one-electron basis functions were combined
to form a set of Slater determinants (as described earlier, the
lowest energy spin state will be projected out by the path
integral procedure) that were then used as the initial
$|\alpha,s>$ for the Monte Carlo simulation. A simulation using
the absolute value of the summand in Eqn.~\ref{Final Result for S}
as the weighting function was performed in which kinks were added,
removed, and changed. An upper limit on the number of kinks
allowed was set to 10 kinks; since the final results were obtained
with 0 or 2 kinks, this did not affect the accuracy of our
results. A list of the states accepted was kept. If the fraction
of configurations that contained more than 0 kinks was greater
than the fraction of configurations that had 0 kinks, the
Hamiltonian was diagonalized using the current list of states and
a new set of N-electron states obtained. These new diagonalized
states were linear combinations of the initial set of Slater
orbitals and thus corresponded to configuration interaction (CI)
wavefunctions. Since the simulation sums over all possible states,
in essence a complete CI calculation is performed. Another
possible Monte Carlo scheme would allow the individual Slater
determinants to be altered during the simulation, which would
correspond to a MCSCF calculation. At most 100 states were
included in the diagonalization to limit the time per
diagonalization. If the set of accepted states exceeded 100 at the
time of diagonalization, only the 100 most prevalent states were
included. Once 5000 iterations had occurred with no
diagonalizations, the run was terminated and these final 5000
energies used to determine energies. At the end of the
calculation, the ground state would correspond to a high quality
CI ground state; if the state does not correspond to the complete
CI wavefunction, then kinks will be added to correct the ground
state. If the adaptive procedure provides the complete CI
wavefunction, then no kinks would ever be introduced, as the
density matrix would be diagonal. In practical calculations, we
expect to stop the adaptive process before the density matrix is
actually diagonal, but when the off-diagonal matrix elements are
so small that the likelihood of adding more than 2 kinks is very
small. In fact, in our calculations, the Monte Carlo procedure
provided such a good estimate of the true ground state that at any
time we found only 0 kinks or 2 kinks (to one of the excited
states).

\section{\label{sec:level4}Application to Atomic Energies}
We have tested this approach by applying it to atomic systems,
using the 6-31G basis set. This set was chosen for its relative
simplicity and reasonable accuracy.  For each atom, He-Ne, each
electron was assigned a specific $s_z $, leading to a fixed total
$S_z $. Thus, the sum in Eqn.~\ref{Qas} used just a single spin
state which was a linear combination of states  $S$ such that $S_z
$ was one of the possible values of $\hat{S}_z $. The initial
basis functions were orthonormalized and the one-particle
Hamiltonian was diagonalized, providing an initial set of states.
As the Monte Carlo simulation was performed and  diagonalizations
proceeded as previously described. Our results are shown in
Table~\ref{ens}, along with the Hartree-Fock and CASSCF energies
from Gaussian 98\cite{g98}. The average sign of the density matrix
demonstrates that the adaptive approach adequately reduces sign
problem to well below what is needed for chemical accuracy.  For
comparison, we note that a shifted-contour auxiliary-field Monte
Carlo calculation of Ne\cite{AFMC2}, using a 4-31G basis set, led
to errors of 0.004 a.u., significantly larger than those found in
the present calculations. For illustrative purposes,
Table~\ref{states} shows the evolution of the coefficients of the
Slater determinants that contribute significantly to the ground
state wavefunction of Be, during the first 4 updates. After the
first 4 updates, only minor changes occurred in the ground state.
It can be seen that the adaptive procedure introduces mixing
between the Slater determinants as needed. The degeneracies seen
can be rationalized on the basis of symmetry.
\begin{table}
\begin{tabular}{|c|c|c|c|c|c|c|}
\hline Atom & E(HF) & E(CASSCF) & E(MC, P=10$^{13}$)&$<Sign>$ &$N_{up}$&$N_{down}$\\
\hline
He &   -2.855160 &-2.8701621 &  -2.8701621(0)&   1.0000(0) &1&1\\
Li &   -7.4312350& -7.4315542 & -7.4315535(6)&   1.0000(0) &2&1\\
Be &  -14.5667641& -14.6135453 & -14.6135468(22)&   1.0000(0)&2&2\\
B &-24.5193448 &  -24.5628917 & -24.5628918(14)&   1.0000(0) &3&2\\
C  &-37.6768656 & -37.7162644 & -37.7162663(24)&   1.0000(0) &4&2\\
N  & -54.3820508& -54.4199396 & -54.4199404(32)&   1.0000(0) &5&2\\
O  & -74.7782342& -74.8394081 & -74.8394091(38) &   0.9992(11) &5&3\\
F  &-99.3602182 & -99.4474231 & -99.4474225(34) &   0.9996(8) &5&4\\
Ne &-128.4738769 & -128.5898023 & -128.5898026(24)&   0.9996(8) &5&5\\
\hline
\end{tabular} \caption{Hartree-Fock (HF), CASSCF, and path
integral (MC) energies (in atomic units), and average sign of the
density matrix for the different atoms studied in this work. The
numbers in parenthesis represent 2 standard deviations. The number
of up- and down-spin electrons is also specified. \label{ens}}
\end{table}

\begin{table}
\begin{tabular}{|c|c|c|c|c|c|c|}
\hline Update Number & E(ground state) & State 1 & State 2&State 3&State 4&State 5\\
\hline &Degeneracy &1&2&1&3&6\\
 \hline
1 &-14.3543&1.0&0.0&0.0&0.0&0.0\\
2 &-14.6053&0.7788&-0.3729&0.1408&-0.1340&-.07652\\
3 &-14.6131&0.7782&-0.3726&0.1407&-0.1339&-.07645\\
4 &-14.6135&0.7715&-0.3767&0.1452&-0.1343&-.07872\\
 \hline
\end{tabular} \caption{Energies (in atomic units) and coefficients of the ground state
wavefunction, as a function of adaptive update, for the first 4
adaptive updates of Be. States are arbitrarily labeled and
correspond to multiple states with degeneracies as indicated. The
Monte Carlo procedure identified all degenerate states and mixed
them with identical coefficients. \label{states}}
\end{table}

\section{\label{level5}Conclusions}
The adaptive, kink-based approach to path integral calculations
has been applied to atomic systems. As was the case in our
previous work, the use of the adaptive approach reduced the sign
problem to a tolerable level. While we have used an adaptive
diagonalization procedure to improve our estimates for the
electronic states, this is not an essential ingredient in the
adaptive approach. For instance, unitary transformations can be
sampled as part of the Monte Carlo process. In addition, we have
not made any simplifying assumptions that will be required when
treating systems with large numbers of basis functions, such as
limiting the type of determinant that can contribute to the ground
state wavefunction. We note that the number of electrons and basis
functions used in this study is on the order of that needed to
study moderately large metal clusters. For example, Na$_{20}$
would require roughly 10 shell-orbitals (orbitals centered at the
origin of the cluster, in accord with the shell model of the
electronic structure) and 20 electrons, if pseudopotentials are
used for the core electrons. Thus, the current method may be
applicable to moderately large systems.

\section{\label{level6}Acknowledgements}
The computers used in these calculations were purchased with funds
from NSF 9977124.

% \appendix

% \section{}
% \label{}

%\begin{thebibliography}{00}


% notes:

% \bibitem{label2}
% If there is a note, it should come last:
% \bibitem{label3} \note
% \end{subbibitems}

%\bibitem{}

%\end{thebibliography}
\bibliographystyle{elsart-num}

\bibliography{rwh}
\end{document}